\documentclass[final,3p,times,twocolumn,sort&compress,10pt]{elsarticle}
\usepackage{lineno,hyperref,amssymb,amsmath}
\usepackage{graphicx,multicol}
\usepackage{subfig}
\usepackage{float}
\usepackage{array}
\newcolumntype{C}[1]{>{\centering\arraybackslash}p{#1}}

\journal{Nuclear Instruments and Methods in Physics Research Section A: Accelerators, Spectrometers, Detectors and Associated Equipment}
\begin{document}

\begin{frontmatter}

\title{Time-of-flight Fourier UCN spectrometer}

\author[Institute1]{G.V. Kulin\corref{mycorrespondingauthor}}
\cortext[mycorrespondingauthor]{Corresponding author}
\ead{kulin@nf.jinr.ru}

\author[Institute1]{A.I. Frank}
\author[Institute1]{S.V. Goryunov}
\author[Institute1]{D.V. Kustov}
\author[Institute2]{P. Geltenbort}
\author[Institute2]{M. Jentschel}
\author[Institute3]{B.Lauss}
\author[Institute3]{Ph. Schmidt-Wellenburg}

\address[Institute1]{Joint Institute for Nuclear Research, Dubna, Russia}
\address[Institute2]{Institut Laue-Langevin, Grenoble, France}
\address[Institute3]{Paul Scherrer Institut, Switzerland}

\begin{abstract}
We describe a new time-of-flight Fourier spectrometer for investigation of UCN diffraction by a moving grating. The device operates in the regime of a discrete set of modulation frequencies. The results of the first experiments show that the spectrometer may be used for obtaining UCN energy spectra in the energy range of 60$\div$200 neV with a resolution of about 5 neV. The accuracy of determination of the line position was estimated to be several units of $10^{-10}$ eV
\end{abstract}


\end{frontmatter}

\section{Introduction}

The UCN spectrometry has been successfully developing during a number of decades. Historically the first method of the UCN spectroscopy was the time-of-flight technique \cite{SteyerlZPhys2521972}. As a direct and universal method of measuring the neutron speed, it has been successfully used in numerous experiments with UCN (see for example \cite{Lynn120b1983,FierlingerNIMA5572006,LavellePhysRevC822010,SharapovPhysRevC882013,MishimaJournPhysConfSer5282014}).

The small energy and large wavelength of UCN have contributed to the appearance of new spectrometry methods, which are unfeasible with thermal neutrons. Specifically, the gravity UCN spectrometry \cite{GroshevPhysLettB341971,KosvintsevNIM1431977,RichardsonNIMA2841989,SchehenhoferPhysRevLett391977,SteyerlZPhysB301978,BondarenkoPhysAtomNucl621999,BondarenkoNIMA4402000,KartashovIntJournNano62007} and the spectrometry based on the use of quantum spectrometric devices (various interference filters) \cite{SteyerlZPhysB301978,BondarenkoPhysAtomNucl621999,SereginJETP731977,PokotilovskiiPTE11980,SteyerlPhysB1511988} have received widespread use.

During the past fifteen years a number of experiments with the gravity UCN spectrometer with interference filters have been performed \cite{FrankPhysAtomNucl662003,FrankPhysLettA3112003,FrankJETPLett782003,FrankJETPLett812005,FrankJETPLett862007}. Due to the narrow transmission spectrum the filters may be used as an effective monochromator and analyzer of energy, and the acceleration of neutrons in the gravitational field of the Earth on their way from the monochromator to the analyzer permits performing energy scanning. The range of energies to be measured is limited in this method by $mgH$, where m is the neutron mass, $g$ is the free fall acceleration and $H$ is the maximum possible distance between both filters. In practice, the energy range did not exceed 20-30 neV. In particular, the effect of neutron energy quantization in UCN diffraction by a moving grating was observed for the first time using such a spectrometer \cite{FrankPhysLettA3112003}. Later this effect was used to test the weak equivalence principle for the neutron \cite{FrankJETPLett862007}.

The continuation of this research \cite{KulinNIMA7922015} and recent theoretical results \cite{BushuevarXivJETP2015} have brought us to the understanding of the need for more detailed investigation of UCN spectra in diffraction by a moving grating. The problem is that the energy range of the resulting spectrum is of the order of 100 neV, which is comparable with the UCN initial energy. At the same time it is desirable that the energy of the spectral lines might be measured with the accuracy of 1$\div$1.5 neV. The gravity UCN spectrometry with interference filters appears to be unsuitable for these purposes.

As it follows from simple estimations the classical TOF method is not suited for this task either, because the ratio of the initial time pulse duration to the pulse repetition period cannot exceed the ratio of the necessary resolution to the width of the time-of-flight spectrum. In our case this ratio is of the order of $10^{-2}$. Such losses in luminosity are absolutely unacceptable.

It is likely that the solution of the problem is the use of any correlation time-of-flight methods. It may be the correlation spectroscopy with pseudo-random flux modulation \cite{CookYahrborough1964,MogilnerPTE21966,SkoldNIM631968,GordonPhysLettA261968,KrooPEPAN81977,CserNIM1841981,FreudenbergNIMA2431986,GutsmiedlPhysB1691991,NovopoltsevInstrExpTech532010} or Fourier spectroscopy with periodic or quasiperiodic flux modulation \cite{HiismakiNineties1985,VirjoNIM731969,ColwellNIM761969,NunesActaCrystA271971,PoyryNIM1261975,PoyryNIM1561978,SchroderJNeutrRes21994,AksenovPhysUsp391996,MaayoufNIMA3981997}.

Having compared the two approaches, we have chosen the Fourier TOF spectroscopy \cite{KulinJINRCommP3201472}. The point is that we had a UCN spectrometer with periodic flux modulation \cite{KulinNIMA7922015}, which could be relatively easily transformed into a TOF Fourier-spectrometer. On the other hand, the use of pseudo-random flux modulation was bound to lead to a dramatic loss of intensity because its realization would severely limit the beam cross-section, which in our case had a shape of a ring.

Such a UCN Fourier spectrometer was built and in the present paper we report the results of its experimental test.

\section{ Time-of-flight Fourier spectroscopy with periodic flux modulation. Idea of the method}

Let the aim of the measurement be the distribution of times $I(t)$ that neutrons spend for the flight along the known path length. This time-of-flight spectrum is evidently related to the velocity distribution. The spectrum might be directly measured if neutrons would be generated in an ideally narrow $\delta$-like time pulse. In the general case the detector response is
\begin{equation}
Z(t) = \int\limits_{0}^{\infty} I(t^\prime) \theta(t-t^\prime)\,\mathrm{d}t^\prime,
\label{eq:generaldetresponse}
\end{equation}
where $\theta(t)$ describes the time dependence of the initial flux.

If the initial flux is modulated harmonically then
\begin{equation}
Z(t) = \int\limits_{0}^{\infty} I(t^\prime) \sin[\omega(t-t^\prime)]\,\mathrm{d}t^\prime,
\label{eq:harmdetresponse}
\end{equation}

Each small part of the time spectrum contributes its harmonic to the detector count rate and
the phase shift of this harmonic relative to the modulation phase is defined by the time of flight and modulation frequency. Since the time spectrum may be represented by a Fourier expansion
\begin{equation}
I(t) = \int\limits_{0}^{\infty} R(\omega) \sin[\omega t-\phi(\omega)]\,\mathrm{d}\omega,
\label{eq:fourierexpan}
\end{equation}
$Z(t)$ in equation \eqref{eq:harmdetresponse} can be interpreted as a Fourier-harmonic of the initial spectrum. For the reconstruction of the initial spectrum in an ideal case it is necessary to find functions $R(\omega)$ and $\phi(\omega)$ in an infinitely large range of frequencies $\omega$.

In modern Fourier spectrometers the correlation between the count rate and delayed function of the beam modulation is measured instead of amplitudes and phases of the count rate oscillation \cite{AksenovPhysUsp391996,MaayoufNIMA3981997}. But since there was a device at our disposal with a fairly stable low-frequency beam modulator of periodic operation, we decided that at the initial stage of the work the measurement with a discrete set of frequencies $\omega_{k}$ would be acceptable. Then the results of the measurement are the amplitudes $R_{k}$ and phases $\phi_{k}$ of the count rate oscillation for all frequencies $\omega_{k}$. Having a rather large amount of such data it is possible to reconstruct the initial time-of-flight spectrum in any time interval
\begin{equation}
I_{exp}(t) = \frac{\pi}{2} \sum_{k} R_{k} \sin(\omega_{k}t+\phi_{k}).
\label{eq:descreterecover}
\end{equation}

Unfortunately, the use of a discrete set of frequencies results in the appearance of phantom fragments in the reconstructed spectrum, which are identical to the real spectrum and repeat with a period $T_{ph} = 2\pi /\Delta \omega$, where $\Delta \omega$ is the step of frequency variation.

In practice, flux modulation is usually realized using the so-called Fourier chopper which consists of a rotating rotor with a large number of slits and a fixed stator with one or several slits. All slits are of equal angular width and separated by areas with the same width that are nontransparent to neutrons. As a result, the flux is modulated by a periodic function $\theta(t)$ in the form of repetitive pulses of triangular shape immediately following each other. This function does not differ too much from a harmonic function, but in addition to the basic frequency $\omega_{k}^{1} = 2\pi /T_{k}$ , there is an admixture of high-order oscillation with frequencies $\omega_{k}^{n} =(2n-1)\omega^{1}_{k}$. Here $T_{k}$ is the pulse repetition period. Only the third-order harmonic (n=2) contributes noticeably to the spectrum, which, if necessary, can be taken into account during processing.

\section{ The spectrometer}
\begin{figure*}[ht]
     \centering
     \minipage{0.32\linewidth}
     \includegraphics[width=\linewidth]{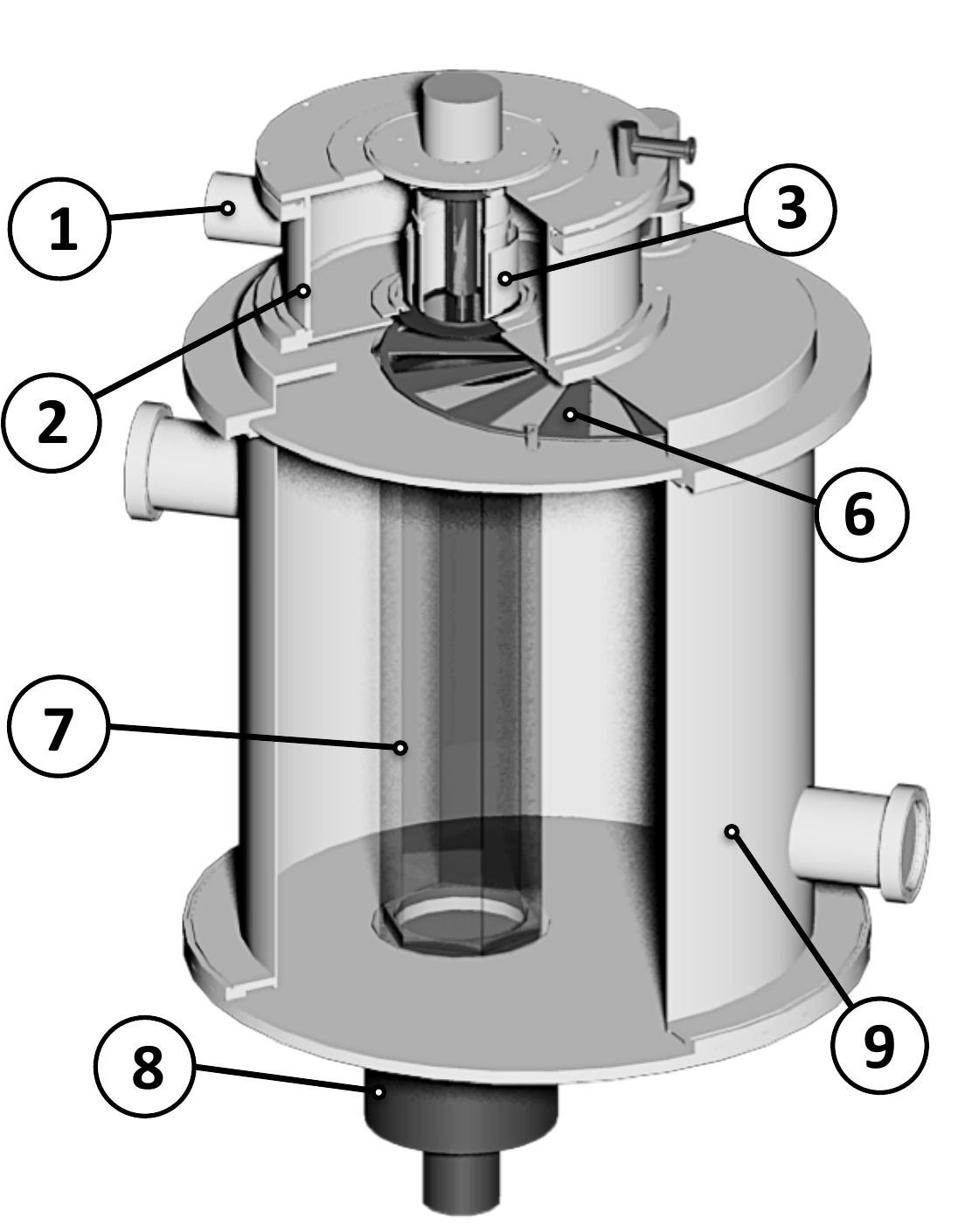}
     \endminipage\hspace{0.1\linewidth}
     \minipage{0.32\linewidth}
     \includegraphics[width=\linewidth]{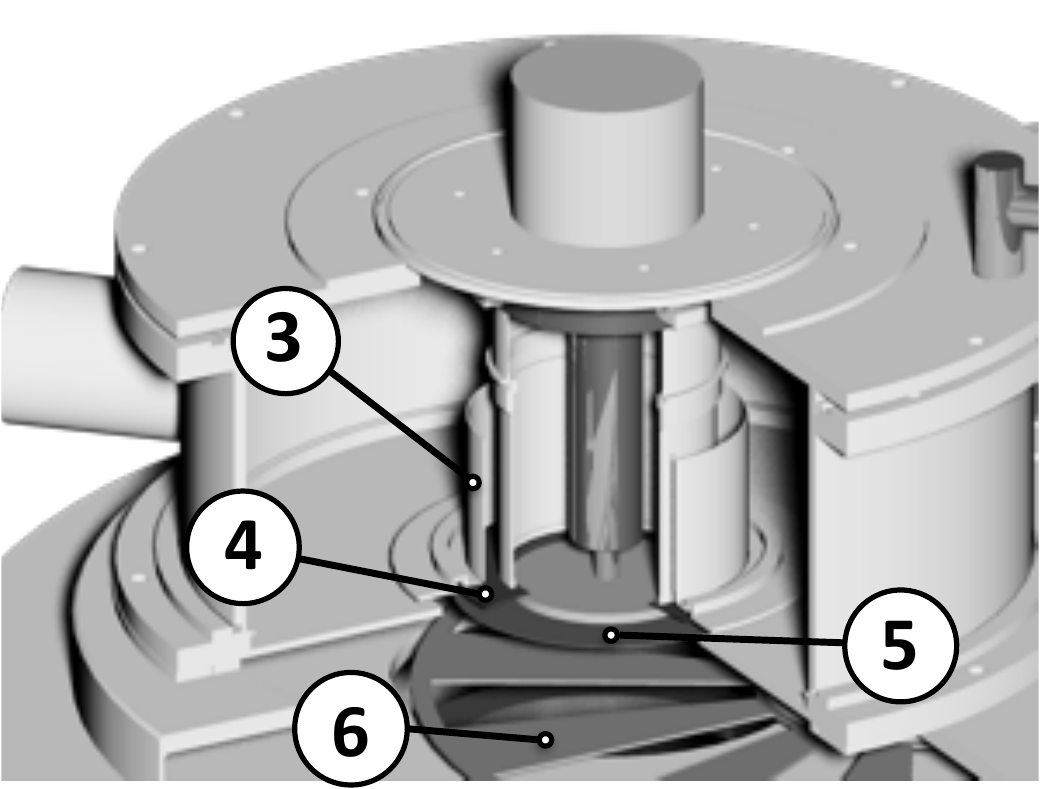}
     \endminipage
    \caption{Time of flight Fourier spectrometer: general view (at the left) and its upper part (at the right): 1 - feeding guide, 2 - entrance chamber, 3 - annular channel, 4 - filter-monochromator, 5- grating, 6- rotor of the Fourier modulator, 7- vertical glass guide, 8- detector, 9- vacuum vessel}
\label{fig:TOFSpectrometer}
\end{figure*}

The spectrometer is a slightly modified version of the device described in \cite{KulinNIMA7922015}. It is shown in Figure\ref{fig:TOFSpectrometer}. Ultracold neutrons are fed to the entrance chamber through the UCN neutron guide and, after a number of reflections, fall down the annular channel with the lower section closed by a monochromator, which is a five-layer Ni-Ti interference filter \cite{BondarenkoPhysAtomNucl621999,BondarenkoNIMA4402000,KulinJINRCommP3201472}. To suppress the background of neutrons with energies higher than the effective potential of nickel, the filter- monochromator is combined with a multilayer "superwindow filter \cite{BondarenkoPhysAtomNucl621999}. In the experiments with a moving grating (see \cite{KulinNIMA7922015} for details) the latter was placed directly below the monochromator. It could be rotated by a motor.

\begin{figure*}[ht]
    \centering
    \minipage{0.4\linewidth}
    \includegraphics[width=\linewidth]{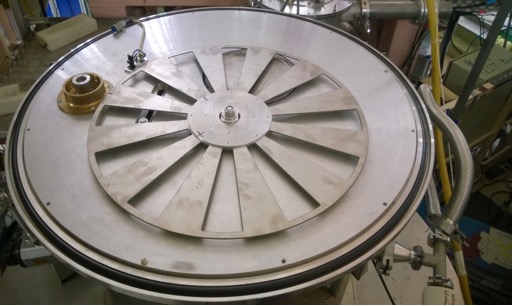}
    \endminipage\hspace{0.01\linewidth}
    \minipage{0.4\linewidth}
    \includegraphics[width=\linewidth]{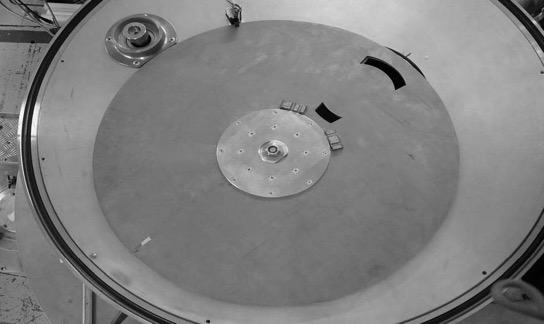}
    \endminipage
    \caption{Rotors for the Fourier chopper (at the left) and for the classical TOF measurement (at the right)}
\label{fig:TOFRotors}
\end{figure*}

The neutron flux with the spectrum which is formed by the combination of two filters and transformed by the quantum modulator -- moving diffraction grating (if the latter is used) comes to the spectrometric part of the device. It comprises a Fourier chopper, vertical neutron guide and detector. The Fourier chopper consists of a rotating rotor and stator. The rotor is a 2-mm-thick titanium disc about 40 cm in diameter with twelve radial slits (see Figure \ref{fig:TOFRotors}). The stator is a titanium diaphragm with only one slit. It is placed at the inlet section of the vertical neutron guide.

As in \cite{KulinNIMA7922015} the rotor is driven by a Phytron stepper motor VSS-65HV located outside the vacuum volume and connected to the rotor via a toothed belt and magnetic coupling. The rotation frequency of the rotor may reach 1800 rpm, which corresponds to a modulation frequency of 360 Hz. For the use of the electronic control system an infrared Honeywell sensor HOA2006-001 and a small slot at the periphery of the rotor are used. Stability of the rotor rotation frequency is of the order of $10^{-4}$.

Upon passing through the modulator, UCNs come to the vertical neutron guide formed by six 95$\times$680 mm flout glass plates and reach the scintillation detector.

\section{Test objects and procedure of the measurement}

To test the spectrometer, four objects were chosen: three interference filters \cite{BondarenkoPhysAtomNucl621999,BondarenkoNIMA4402000,KulinJINRCommP3201472} (hereinafter referred to as filter 1,2,3) and diffraction grating. Their short descriptions are given below.
\begin{enumerate}
\item Filter 1 is a five-layer filter with alternating layers of NiMo and TiZr with design values of thickness 24.5-23.0-48.0-23.0-24.5 nm.
\item Filter 2 is a nine- layer filter with alternating layers of NiMo and TiZr with design values of thickness 12.5-23-32-23-38-23-32 -23-12.5 nm.
\item Filter 3 is a five- layer filter with alternating layers of Ni(N) and TiZr with design values of thickness 23-23.7-24.5-23.7-23.0 nm.
\item Rotating diffraction grating. It was prepared on the surface of a silicon disc 150 mm in diameter and 0.6 mm thick. Radial grooves were made in the peripheral region of the disc, which is a ring with an average diameter of 12 cm and a width of about 2 cm. The widths of the grooves are proportional to the radius and this proportionality ensures a constant angular distance between the grooves equal to a half period. The angular period of the structure is exactly known to be $\alpha=2\pi /N$ with N = 94500. The grating was manufactured by Qudos Technology Ltd. \footnote{Qudos Technology Ltd, Rutherford Appleton Laboratory, OX11 0QX Chilton, United Kingdom}
\end{enumerate}

 \begin{figure}[h]
	\includegraphics[width=1.0\linewidth]{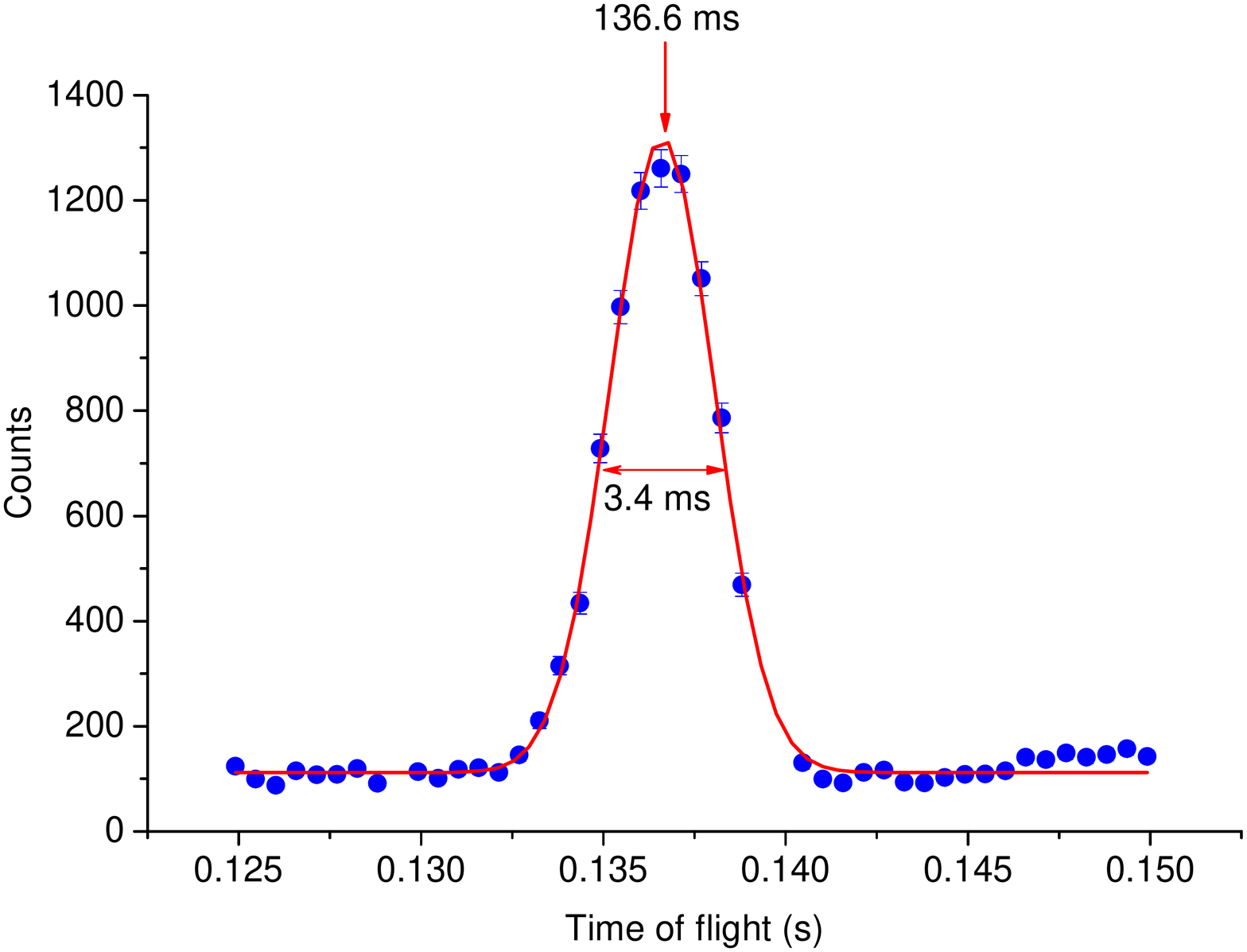}
	\caption{Time-of-flight spectrum formed by filter 1 and "superwindow"}
	\label{fig:filter1}
\end{figure}
\begin{figure}[h]
	\includegraphics[width=1.0\linewidth]{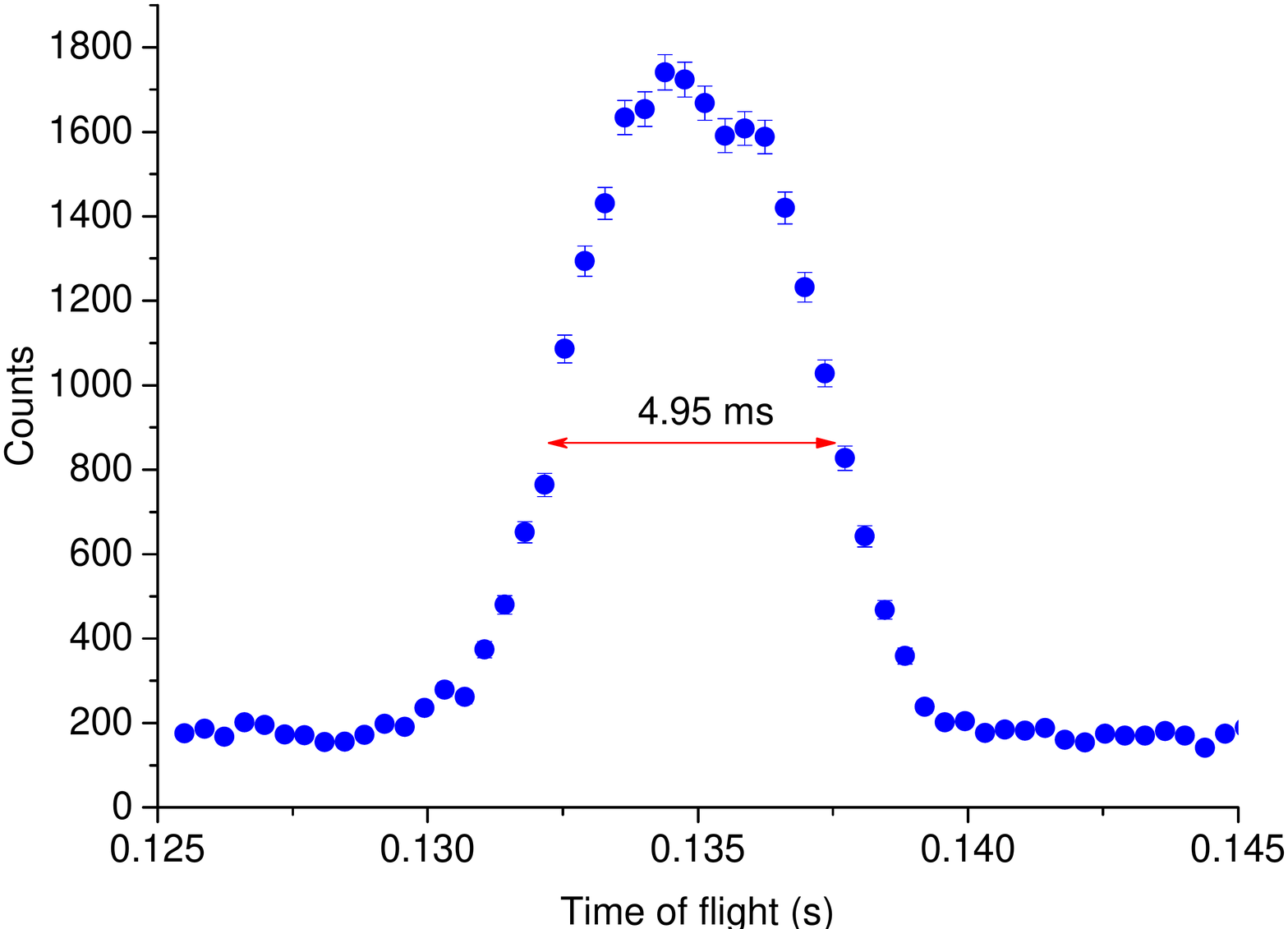}
	\caption{Time-of-flight spectrum formed by filter 2 and "superwindow"}
	\label{fig:filter2}
\end{figure}

 Herein, we do not provide the calculated transmittivity of filters 1 and 2, as it is very similar to that displayed in Figures (3) and (4) of \cite{KulinNIMA7922015}. Spectra of UCN transmitted through these filters were obtained by a standard TOF method using the same spectrometer as in \cite{KulinNIMA7922015}. The chopper used in the measurement is shown in Figure \ref{fig:TOFRotors}.

Filter 3 was characterized by the splitting of the transmission line (see details in \cite{BondarenkoPhysAtomNucl621999,BondarenkoNIMA4402000,FrankProcSPIE37671999}). In the regime of Fourier spectrometry filters 1-3 together with the filter-superwindow served as a monochromator. In the experiments with the moving grating the latter operated as a high-frequency phase modulator, which resulted in the splitting of the spectrum formed by the monochromator into a number of lines \cite{FrankPhysLettA3112003,FrankJETPLett812005}].

As noted above, when the setup was used as a Fourier spectrometer the neutron flux was modulated by the Fourier chopper operating with the specified frequency. The measurement system registered the time of arrival of pulses from the sensor of the chopper and from the detector and these data were recorded sequentially to a file. The modulation frequency $f_{k}$ ranged from 6 to 360 Hz. It was increased with a step of 6 Hz (in some measurements by 12 Hz) and upon reaching the maximum value decreased again. The duration of each measurement was usually 1000 s. The count rate was several counts per second.

\begin{figure}[h]
	\centering
	\includegraphics[width=1.0\linewidth]{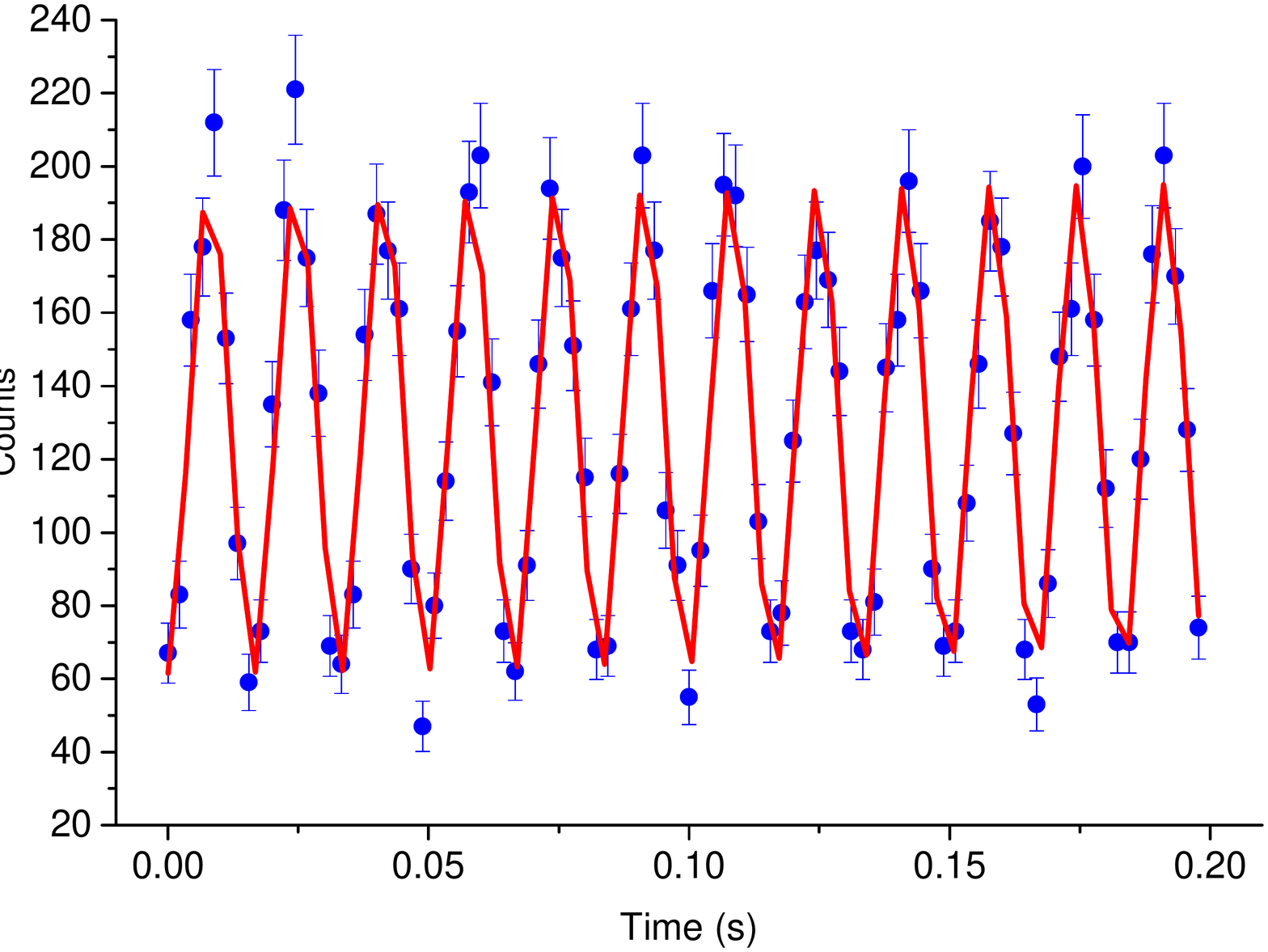}
	\caption{Count rate oscillation at the measurement with filter 1. Modulation frequency f=60 Hz}
	\label{fig:countrateoscill}
\end{figure}
\begin{figure}[h]
	\centering
	\includegraphics[width=1.0\linewidth]{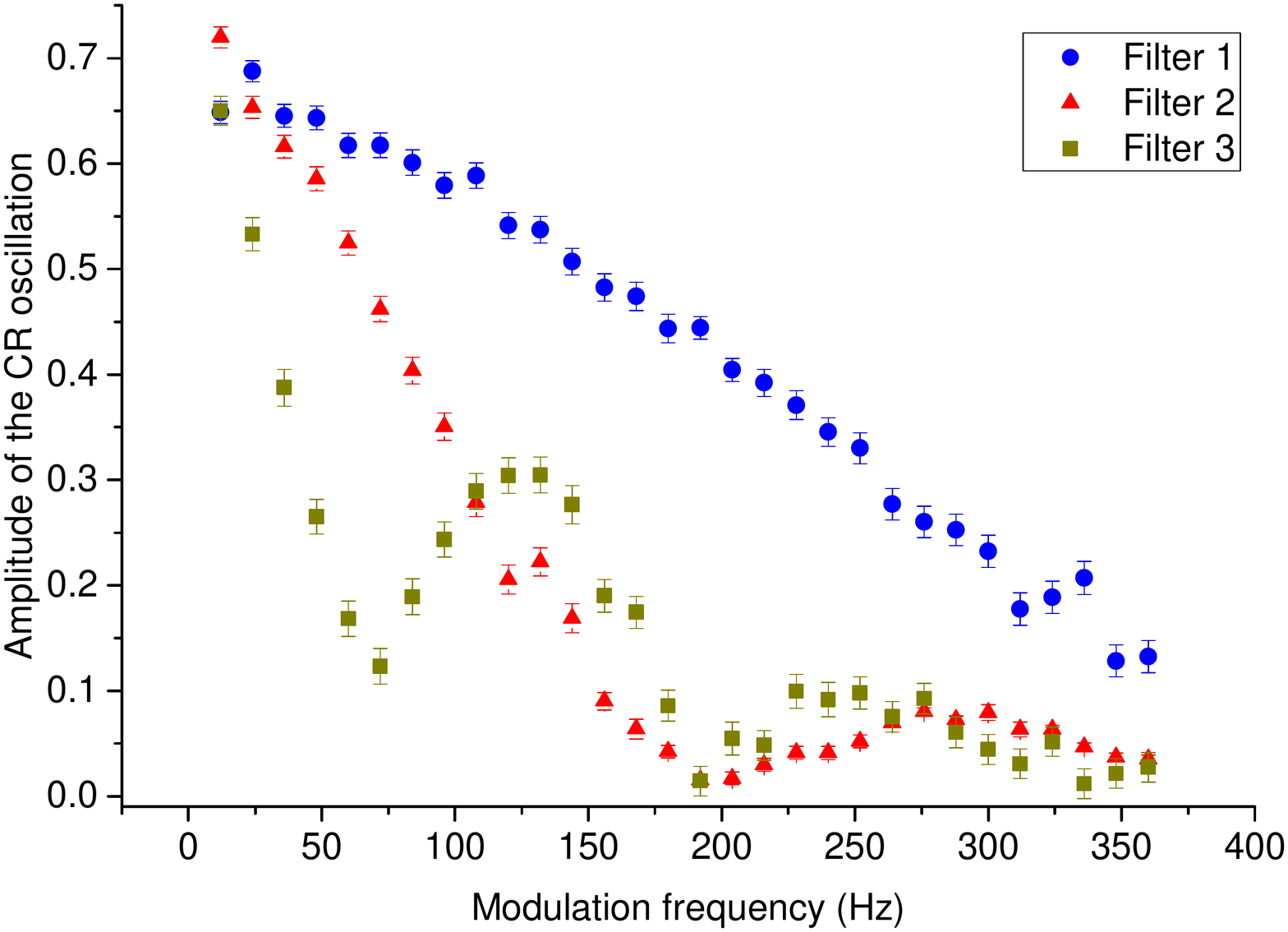}
	\caption{The dependence of the amplitude of the count rate oscillation on the modulation frequency for the measurements with three different filters}
	\label{fig:ampdependence}
\end{figure}

The data obtained in each measurement with the modulation frequency $f_{k}$ were fitted by the sine function, as shown in Figure \ref{fig:countrateoscill}. The result of such fitting was the values of amplitude $R_{k}$ and phase $\phi_{k}$ of the count rate oscillation. The dependence of the amplitude on the modulation frequency is shown in Figure 6. After substitution of the obtained  $R_{k}$ and phase $\phi_{k}$ into the equation \eqref{eq:descreterecover} and taking into account that  $\omega_{k}=2 \pi f_{k}$, it is easy to obtain the time-of-flight spectrum.

It should be noted that due to the vertical orientation of the spectrometer and the effect of the Earth's gravity the neutron time of flight does not linearly depend on the initial velocity. Sometimes for the correct interpretation of the results it was necessary to recalculate the time spectrum into the initial energy spectrum. In addition to the relation of energy and time of flight
given by the formula
\begin{equation}
E = \frac{m}{2} \left( \frac{ H^{2}}{t^{2}}-gH+\frac{g^{2}t^{2}}{4} \right) ,
\label{eq:energyTOFrelation}
\end{equation}
it was also necessary to take into account the nonlinear relation between the widths of time and energy "channels" on the abscissa axis. The relation between $N_{t}$ and $N_{E}$ values, which are proportional to the intensities in time and energy channels, respectively, is given by the relation
\begin{equation}
N_{E} = N_{t} \left[ m \left(\frac{H^{2}}{t^{3}}-\frac{g^{2}t}{4}\right) \right]^{-1},
\label{eq:channelsrelation}
\end{equation}
where $m$ is the neutron mass, $g$ is the free fall acceleration and $H$ is the height of the Fourier chopper above the detector.

\section{Experimental results}

\begin{figure}[H]
	\centering
	\minipage{1.0\linewidth}
	\includegraphics[width=\linewidth]{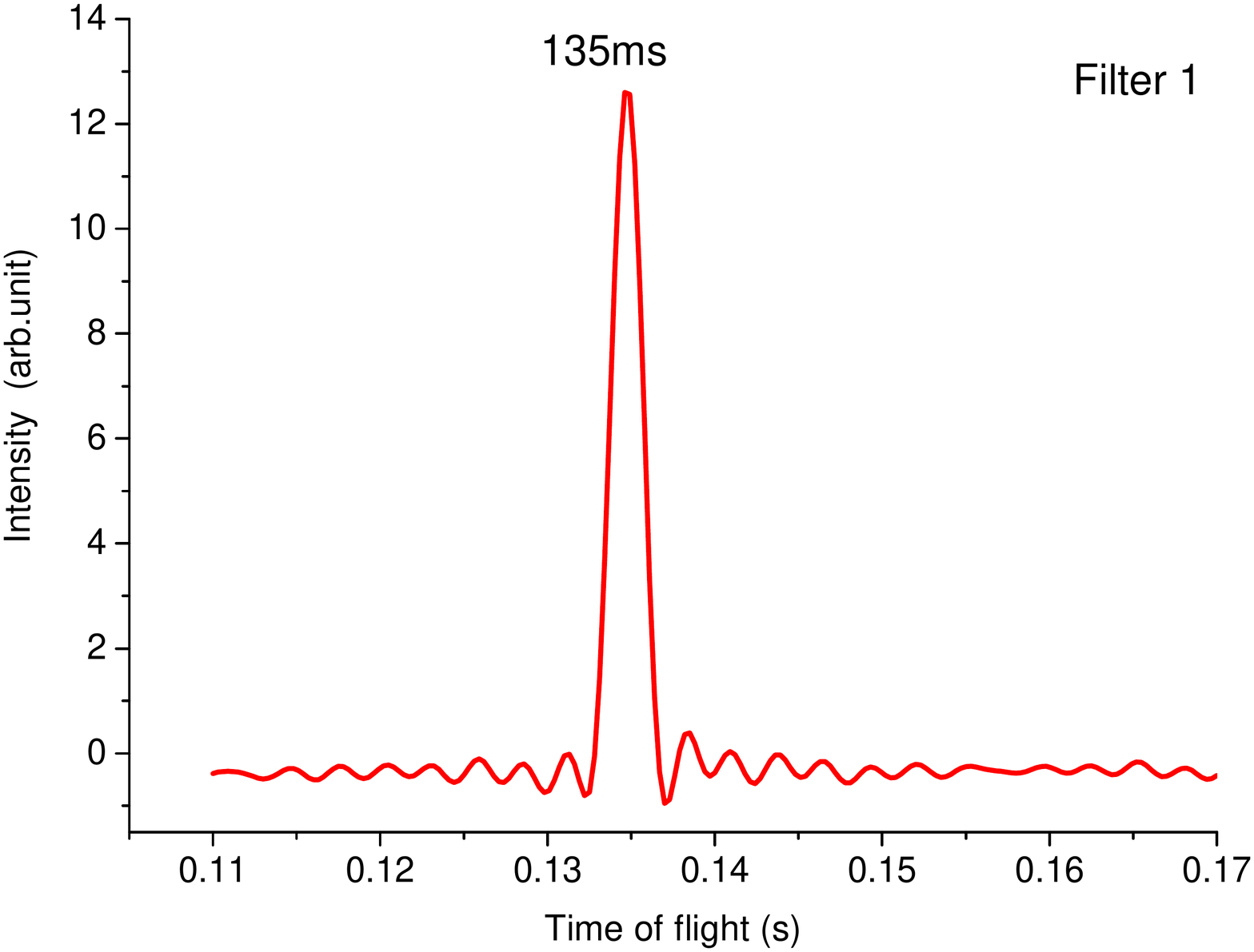}
    \endminipage\hfill
    \minipage{1.0\linewidth}
	\includegraphics[width=\linewidth]{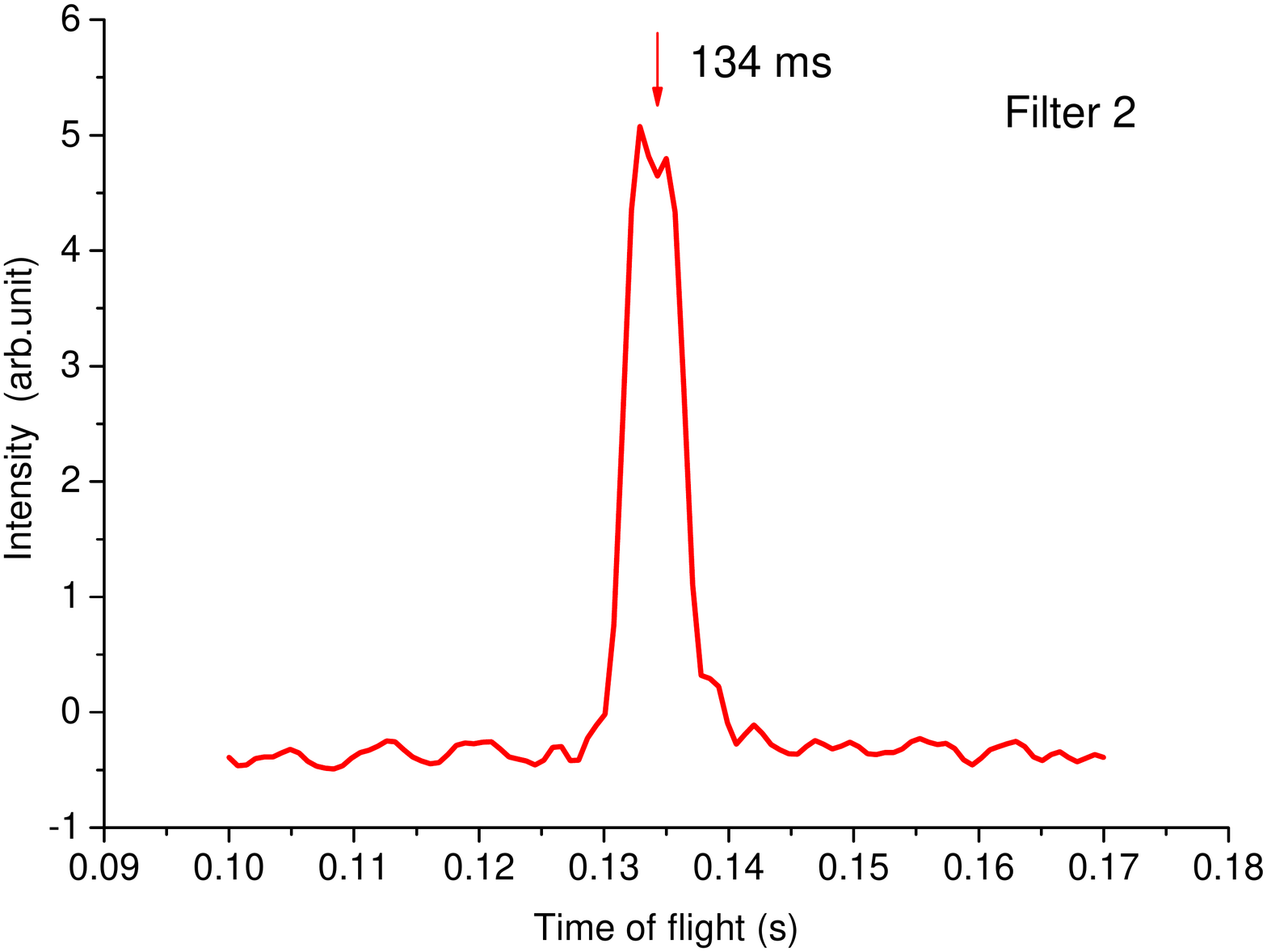}
    \endminipage\hfill
    \minipage{1.0\linewidth}
	\includegraphics[width=\linewidth]{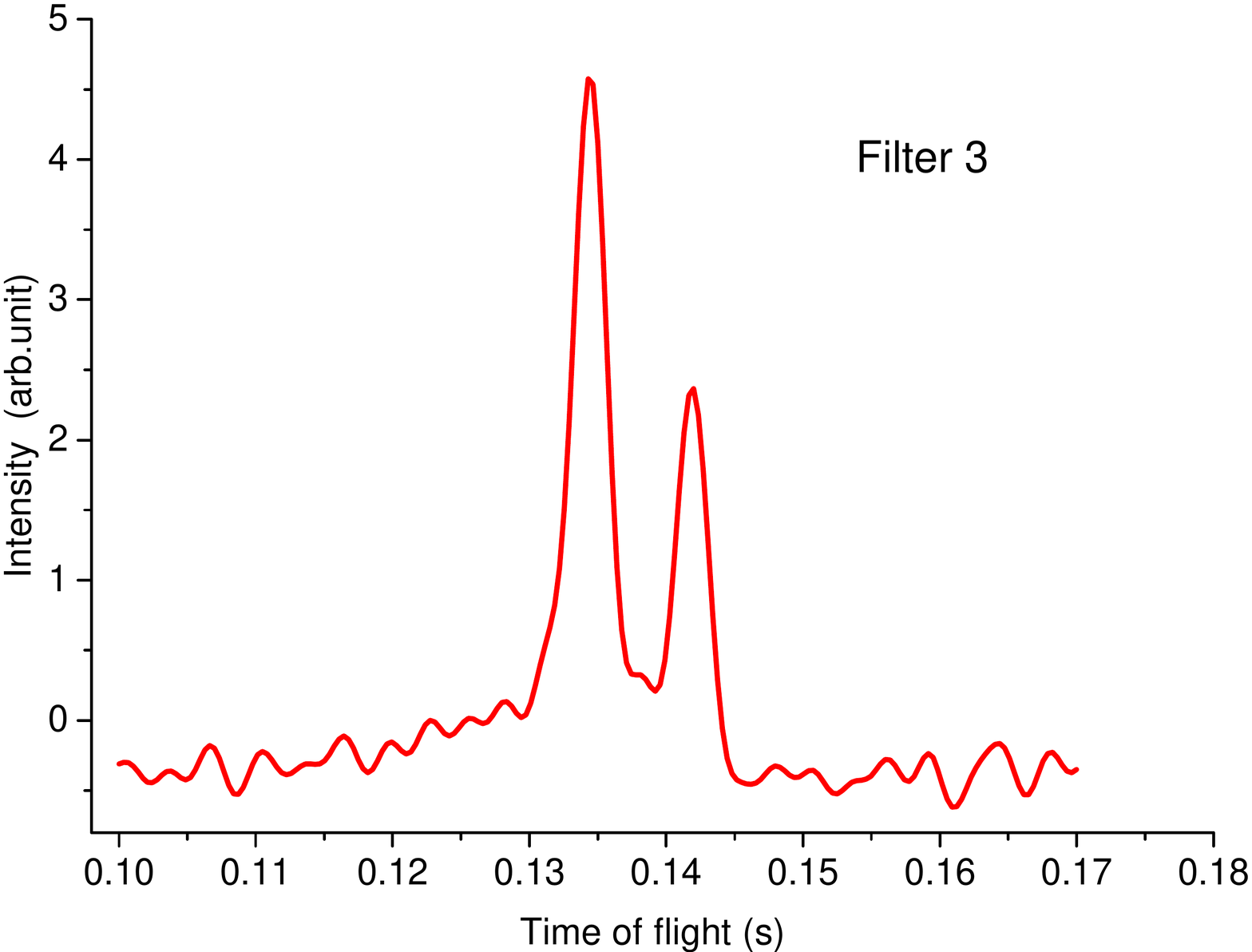}
	\endminipage\hfill
	\caption{Time-of-flight spectra measured by the Fourier spectrometer. Energy spectra formed by filters 1-3 in combination with filter "superwindow"}
	\label{fig:NIFFourierTOFSpectra}
\end{figure}

The time-of-flight spectra obtained as a result of Fourier synthesis are shown in Figure \ref{fig:NIFFourierTOFSpectra}. It is interesting to compare the positions of the peaks in the TOF spectra formed by filter 1 and obtained using two different methods (see Figure \ref{fig:filter1} and the upper plot in Figure \ref{fig:NIFFourierTOFSpectra}). They are rather similar and the small difference of the order of 1.5 ms is probably due to a small error in the time delay between the master-pulse and the moment of opening of the chopper window.

As mentioned above, the third filter was distinguished by the splitting of the energy line in the spectrum of transmitted neutrons. Such spectrum has never been measured by the classical TOF method, but we had the results of the measurement of the convolution of the filter transmission function with the transmission function of another filter-analyzer (so-called scanning curve). The measurement was made in 2000 using the Gravity UCN spectrometer \cite{BondarenkoPhysAtomNucl621999, BondarenkoNIMA4402000, FrankProcSPIE37671999}. Knowing the convolution of two transmission functions it is possible to determine the difference between the energies of two lines with a high precision, but not their absolute values. That is because the parameters of the filter-analyzer were known only from calculations and the accuracy of these estimates was of the order of some neV.

\begin{figure}[H]
    \centering
    \includegraphics[width=1.0\linewidth]{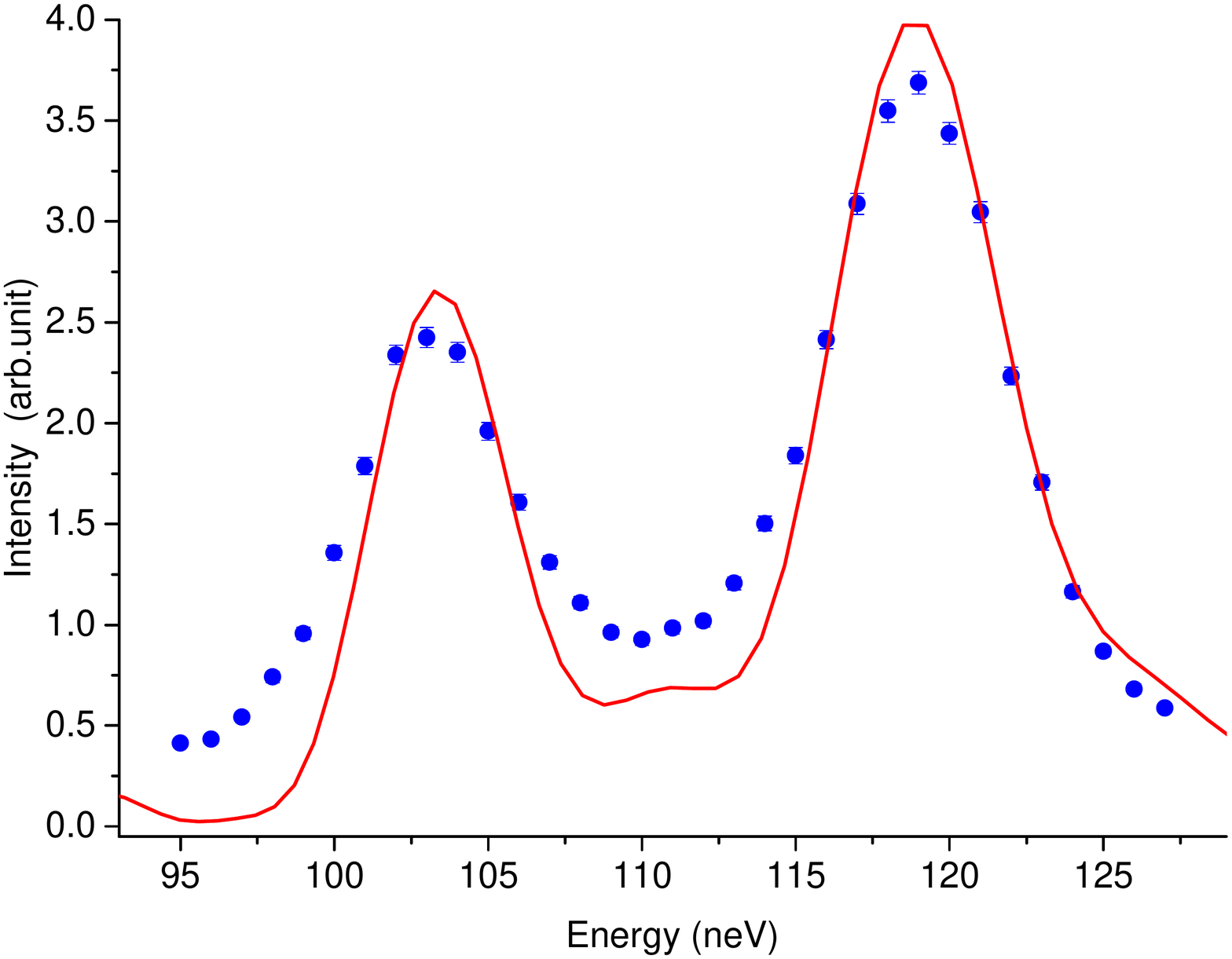}
    \caption{Spectra of neutrons transmitted through filter 3 obtained by two different methods. The solid line corresponds to the spectrum measured by the Fourier spectrometer. The dots are the scanning curve measured in 2000 by the UCN Gravity spectrometer. See the text for details}
\label{fig:filter3spectrum}
\end{figure}

To compare the spectrum measured with filter 3 (see the lower plot in Figure \ref{fig:NIFFourierTOFSpectra}) with the previous results, the obtained TOF spectrum was recalculated to the energy spectrum in accordance with formulas \eqref{eq:energyTOFrelation}, \eqref{eq:channelsrelation}. To match two spectra, the energy scale in the plot of the results of 2000 was shifted by 4 neV. This value is likely to be the estimation of the error of the calculated position of the used filter-analyzer. The results of this comparison are shown in Figure \ref{fig:filter3spectrum}. For illustrative purposes the plot of TOF spectrum was normalized to the scanning curve. 

For a qualitative estimation of measurement precision we fitted the data that had been obtained using the gravity spectrometer by two Gaussians with a linearly increasing background. The distance between two peaks was found as 15.15 $\pm$ 0.26 neV. For the spectrum obtained as a result of Fourier synthesis such a procedure is not quite correct, as it is formed not by a set of points with error bars, but is a continuous curve. Nevertheless, the positions of peak centers may be found by fitting the curve by Gaussians. The distance between the peaks obtained in such a way was 15.5 neV. The error of this value was not defined due to uncertainty in the errors of the fitted spectrum. However, comparison of these two values allows to estimate the determination accuracy for the peak position to be several units of $10^{-10}$ neV.

\begin{figure}[H]
	\includegraphics[width=1.0\linewidth]{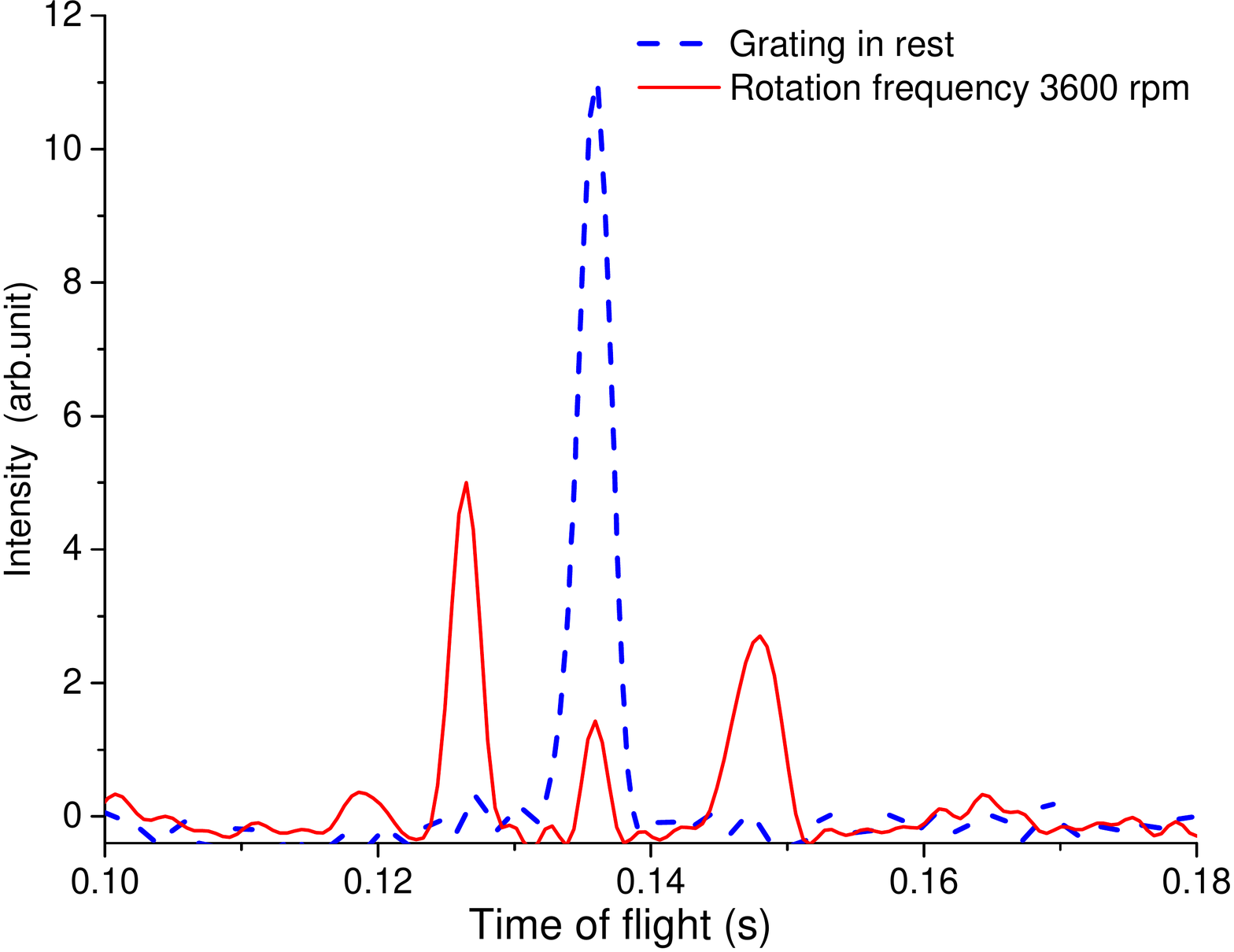}
	\caption{Time-of-flight spectrum of UCN passing through the grating in rest (dashed line) and rotating grating (solid line)}
	\label{fig:gratingTOFspectrum}
\end{figure}

\begin{figure}[H]
	\includegraphics[width=1.0\linewidth]{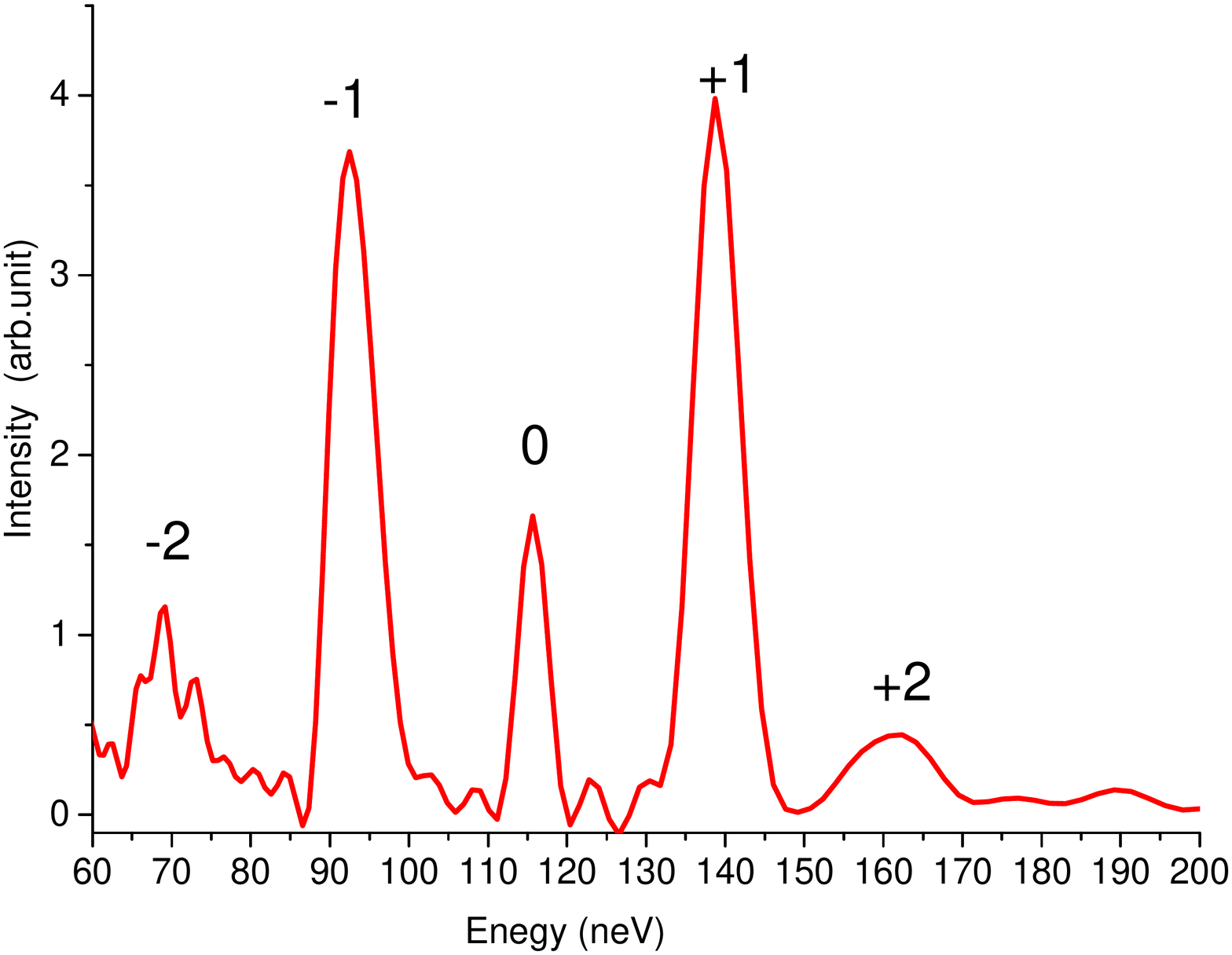}
	\caption{Energy spectrum of UCN passing through the grating rotating with 3600 rpm. Peaks of the zero, $\pm$ first and $\pm$ second diffraction orders can be seen}
	\label{fig:gratingEspectrum}
\end{figure}

As noted above, our aim was to create a setup appropriate for obtaining UCN spectra in a rather wide energy range. The first attempt to perform such a measurement has been done and the time-of-flight spectra of neutrons passing through the diffraction grating in rest and being rotated have been measured in the energy range from 60 to approximately 200 nev (see Figures \ref{fig:gratingTOFspectrum}, \ref{fig:gratingEspectrum}). Although these first measurements have been done with rather pure statistics the lines of the zero, $\pm$ first and $\pm$ second diffraction orders can be clearly seen. A more detailed discussion of
these results and comparison with the theory \cite{BushuevarXivJETP2015} will be reported elsewhere.


The zero-order peak that can be seen in both figures corresponds to the neutrons whose energy has not changed when passing through the grating. 
The position of the peak was measured three times: when the grating was 
at rest and when it rotated at two different speeds. Since the three measurements were quite independent, their results may be used to assess the accuracy of the measurement of the energy position of lines in the spectrum. To compare the results, the lines were fitted by a Gaussian function. The obtained results are shown in the table below.

\begin{table}[h]
	\caption {Parameters of the zero-order diffraction peak obtained in different measurements} \label{tab:zeororder} 
	\centering
	\small
	\begin{tabular}{C{2.5cm}C{2cm}C{2cm}}
		\hline
		Rotation speed of the grating (rpm) & Position of the maximum, neV & FWHM (neV)  \\ \\ \hline
		0 & 115.78 $\pm$ 0.051 & 5.98 $\pm$ 0.134 \\
		3600 & 115.68 $\pm$ 0.02 & 4.49 $\pm$ 0.092  \\
		4800 & 115.90 $\pm$ 0.06 & 4.33 $\pm$ 0.21  \\ \hline
	\end{tabular}
\end{table}

It should be stressed that the errors presented in the Table above cannot be interpreted as measurement errors. They are the errors from fitting the continuous spectra of unknown accuracy by Gaussians. The meaningful data are the position of the peak center of gravity and FWHM.

It can be seen that the obtained positions of the zero-order line maxima differ by $(1\div2)\times10^{-10}$ eV, which corresponds to the error in the time of flight of the order of 0.1 ms. These data qualitatively characterize the spectrometric quality of the device. As for the resolution of the spectrometer it corresponds to the FWHM of peaks. As can be seen from the table it is of the order of 5 neV, which corresponds to the time resolution of the order of 3 ms. It is close to the theoretical estimate, which is equal to the minimum modulation period of a Fourier chopper, which in our case was 2.8 ms.

\section{Conclusion}
We reported on the development of the time-of-flight UCN Fourier spectrometer designed mainly for studying the non-stationary phenomenon of neutron diffraction by a moving grating. This device is a modification of the gravity spectrometer \cite{KulinNIMA7922015} and that is why the first tests of the new device were carried out in the regime of the discrete spectrum of beam modulation and off- line data processing. This regime is not optimal but it made it possible to obtain the first results in a relatively short time.
The results of the test have shown that this device allows UCN energy spectra to be measured in the energy range of 60$\div$200 neV with the resolution of the order of 5 neV. The accuracy of the determination of the line position was estimated was estimated to be several units of $10^{-10}$ eV.

The activities on further improvement of the device will be aimed at increasing its luminosity. In addition, the spectrometer will be modified for using a continuous spectrum of modulation frequencies, which will improve the parameters of the device.

The authors are very grateful to Tomas Brenner for his outstanding technical support. This work was supported by the Russian Foundation for Basic Research (RFBR grant 15-02-02509).


\bibliographystyle{elsarticle-num} 
\bibliography{references}

\begin{thebibliography}{10}
\expandafter\ifx\csname url\endcsname\relax
  \def\url#1{\texttt{#1}}\fi
\expandafter\ifx\csname urlprefix\endcsname\relax\def\urlprefix{URL }\fi
\expandafter\ifx\csname href\endcsname\relax
  \def\href#1#2{#2} \def\path#1{#1}\fi

\bibitem{SteyerlZPhys2521972}
A.~Steyerl, Zeitschrift f\"{u}r Physik 252 (1983) 371.

\bibitem{Lynn120b1983}
J.~Lynn, W.~Miller, T.~Dombeck, Physica 120b (1983) 114.

\bibitem{FierlingerNIMA5572006}
P.~Fierlinger, A.~Pichlmaier, H.~Rauch, Nuclear Instruments and Methods in
  Physics Research Section A: Accelerators, Spectrometers, Detectors and
  Associated Equipment 557 (2006) 572.

\bibitem{LavellePhysRevC822010}
C.~M. Lavelle, C.-Y. Liu, W.~Fox, et~al., Physical Review C 82 (2010) 015502.

\bibitem{SharapovPhysRevC882013}
E.~I. Sharapov, C.~L. Morris, M.~Makela, Physical Review C 88 (2013) 037601.

\bibitem{MishimaJournPhysConfSer5282014}
K.~Mishima, S.~Imajo, M.~Hino, et~al., Journal of Physics: Conference Series
  528 (2014) 012030.

\bibitem{GroshevPhysLettB341971}
L.~Y. Groshev, V.~N. Dvoretsky, A.~M. Demidov, et~al., Physics Letters B 34
  (1971) 293.

\bibitem{KosvintsevNIM1431977}
Y.~Y. Kosvintsev, E.~N. Kulagin, Y.~A. Kushnir, et~al., Nuclear Instruments and
  Methods 143 (1977) 133.

\bibitem{RichardsonNIMA2841989}
D.~J. Richardson, S.~K. Lamoreaux, Nuclear Instruments and Methods in Physics
  Research Section A: Accelerators, Spectrometers, Detectors and Associated
  Equipment 284 (1989) 192.

\bibitem{SchehenhoferPhysRevLett391977}
H.~Schehenhofer, A.~Steyerl, Physical Review Letters 39 (1977) 1310.

\bibitem{SteyerlZPhysB301978}
A.~Steyerl, Zeitschrift f\"{u}r Physik B Condensed Matter 30 (1978) 235.

\bibitem{BondarenkoPhysAtomNucl621999}
I.~Bondarenko, V.~I. Bodnarchuk, S.~N. Balashov, Physics of Atomic Nuclei 62
  (1999) 721.

\bibitem{BondarenkoNIMA4402000}
I.~Bondarenko, S.~Balashov, A.Cimmino, et~al., Nuclear Instruments and Methods
  in Physics Research Section A: Accelerators, Spectrometers, Detectors and
  Associated Equipment 440 (2000) 591.

\bibitem{KartashovIntJournNano62007}
D.~Kartashov, E.~Lychagin, A.~Muzychka, et~al., International Journal of
  Nanoscience 6 (2007) 501.

\bibitem{SereginJETP731977}
A.~Seregin, Journal of Experimental and Theoretical Physics 73 (1977) 1634.

\bibitem{PokotilovskiiPTE11980}
Y.~Pokotilovskii, A.~A. Stoyka, I.~G. Shelkova, Pribory i Tekhnika Eksperimenta
  1 (1980) 62, (in Russian).

\bibitem{SteyerlPhysB1511988}
A.~Steyerl, W.~Drexel, S.~Malik, E.~Gutsmiedl, Physica B 151 (1988) 36.

\bibitem{FrankPhysAtomNucl662003}
A.~I. Frank, V.~I.Bodnarchuk, P.~Geltenbort, et~al., Physics of Atomic Nuclei
  66 (2003) 1831.

\bibitem{FrankPhysLettA3112003}
A.~Frank, S.~Balashov, I.~Bondarenko, et~al., Physics Letters A 311 (2003) 6.

\bibitem{FrankJETPLett782003}
A.~I. Frank, P.~Geltenbort, G.~V. Kulin, A.~N. Strepetov, Journal of
  Experimental and Theoretical Physics Letters 78 (2003) 188.

\bibitem{FrankJETPLett812005}
A.~I. Frank, P.~Geltenbort, G.~V. Kulin, et~al., Journal of Experimental and
  Theoretical Physics Letters 81 (2005) 427.

\bibitem{FrankJETPLett862007}
A.~I. Frank, P.~Geltenbort, M.~Jentschel, et~al., Journal of Experimental and
  Theoretical Physics Letters 86 (2007) 225.

\bibitem{KulinNIMA7922015}
G.~Kulin, A.I.Frank, S.V.Goryunov, et~al., Nuclear Instruments and Methods in
  Physics Research Section A: Accelerators, Spectrometers, Detectors and
  Associated Equipment 792 (2015) 38.

\bibitem{BushuevarXivJETP2015}
V.~Bushuev, A.~Frank, G.~Kulin, Journal of Experimental and Theoretical Physics
  Letters 148~(5), (in press).
\newblock \href {http://arxiv.org/abs/1502.04751v1}
  {\path{arXiv:1502.04751v1}}.

\bibitem{CookYahrborough1964}
H.~Cook-Yahrborough, in: Instrumentation Techniques in Nuclear Pulse Analysis,
  Proceedings of a Conference Held at the U.S. Naval Postgraduate School,
  National Academy of Science--National Research Council, Washington DC, 1964,
  p. 207.

\bibitem{MogilnerPTE21966}
A.~Mogilner, O.~A. Salnikov, L.~Timochin, Pribory i Tekhnika Eksperimenta 2
  279, (in Russian).

\bibitem{SkoldNIM631968}
K.~S\"{o}ld, Nuclear Instruments and Methods 63 (1968) 114.

\bibitem{GordonPhysLettA261968}
J.~Gordon, N.~Kro\'{o}, G.~Orban, et~al., Physics Letters A 26 (1968) 122.

\bibitem{KrooPEPAN81977}
N.~Kro\'{o}, L.~Cher, Physics of Elementary Particles and Atomic Nuclei 8
  (1977) 1412.

\bibitem{CserNIM1841981}
L.~Cser, F.~Ferenczy, N.~Kro\'{o}, G.~Rubin, et~al., Nuclear Instruments and
  Methods 184 (1981) 431.

\bibitem{FreudenbergNIMA2431986}
U.~Freudenberg, W.~Glaser, Nuclear Instruments and Methods in Physics Research
  Section A: Accelerators, Spectrometers, Detectors and Associated Equipment
  243 (1986) 429.

\bibitem{GutsmiedlPhysB1691991}
E.~Gutsmiedl, R.~Golub, J.Butterworth, Physica B: Condensed Matter 169 (1991)
  503.

\bibitem{NovopoltsevInstrExpTech532010}
M.~I. Novopoltsev, Y.~N. Pokotilovski, Instruments and Experimental Techniques
  53 (2010) 635.
\newblock \href {http://arxiv.org/abs/1008.1419v1} {\path{arXiv:1008.1419v1}}.

\bibitem{HiismakiNineties1985}
P.~Hiism\"{a}ki, V.A.Trunov, O.~Antson, et~al., in: Neutron scattering in the
  'Nineties, Proceedings of a Conference on Neutron Scattering in the
  'Nineties, International Atomic Energy Agency, Vienna, 1985, p. 435.

\bibitem{VirjoNIM731969}
A.Virjo, Nuclear Instruments and Methods 73 (1969) 189.

\bibitem{ColwellNIM761969}
J.~Colwell, S.~Lehinan, J.~P.H.~Miller, W.~Whittemore, Nuclear Instruments and
  Methods 76 (1969) 135.

\bibitem{NunesActaCrystA271971}
A.~Nunes, R.~Natans, B.~Schoenborn, Acta Crystallographica A27 (1971) 284.

\bibitem{PoyryNIM1261975}
H.~P\"{o}yry, P.~Hiism\"{a}ki, A.~Virjo, Nuclear Instruments and Methods 126
  (1975) 421.

\bibitem{PoyryNIM1561978}
H.~P\"{o}yry, Nuclear Instruments and Methods 156 (1978) 499.

\bibitem{SchroderJNeutrRes21994}
J.~Schr\"{o}der, V.~A. Kudryashev, J.~M. Keuter, et~al., Journal of Neutron
  Research 2 (1994) 129.

\bibitem{AksenovPhysUsp391996}
V.~Aksenov, A.~Balagurov, Physics-Uspekhi 39 (1996) 897.

\bibitem{MaayoufNIMA3981997}
R.~Maayouf, I.~Abdel-Latif, A.~El-Kady, A.~El-Shafey, Y.~E.-S. M.~Khalil,
  Nuclear Instruments and Methods in Physics Research Section A: Accelerators,
  Spectrometers, Detectors and Associated Equipment 398 (1997) 295.

\bibitem{KulinJINRCommP3201472}
G.~Kulin, D.~Kustov, A.I.Frank, et~al., JINR Communication, Dubna, (in Russian)
  (2014).

\bibitem{FrankProcSPIE37671999}
A.~I.Frank, S.~Balashov, V.~I.Bodnarchuk, et~al., in: EUV, X-Ray, and Neutron
  Optics and Sources, Vol. 3767 of Proceedings of SPIE, SPIE, Denver, CO, USA,
  1999, p. 360.

\end{thebibliography}


\begin{thebibliography}{}
\expandafter\ifx\csname url\endcsname\relax
  \def\url#1{\texttt{#1}}\fi
\expandafter\ifx\csname urlprefix\endcsname\relax\def\urlprefix{URL }\fi
\expandafter\ifx\csname href\endcsname\relax
  \def\href#1#2{#2} \def\path#1{#1}\fi

\end{thebibliography}

\end{document}